\documentclass[twocolumn,showpacs,prd,floatfix,axodraw]{revtex4}
\usepackage{mathrsfs}
\usepackage{graphicx,,booktabs,bm}
\usepackage{overpic}
\usepackage{color}
\usepackage{amssymb}

\usepackage{mathrsfs,bm,amsmath,amssymb}
\usepackage{longtable,lscape}
\usepackage{txfonts}
\usepackage{amssymb}
\usepackage{indentfirst}
\usepackage{graphicx,,booktabs}
\usepackage{multirow,ulem}

\usepackage{color}
\usepackage{amssymb}

\definecolor{cover}{rgb}{0.77,0.87,0.88}
\definecolor{blueone}{rgb}{0.1,0.1,.7}
\definecolor{citec}{rgb}{0.14,0.47,0.09}
\definecolor{two}{rgb}{0.0,0.5,0.}
\definecolor{three}{rgb}{.5,.1,0.15}
\usepackage[bookmarks=true,bookmarksopen=false,plainpages=false,breaklinks=true,
   bookmarksnumbered=true,hypertexnames=false,
   filecolor=blue,urlcolor=three,menucolor=three,
   linkcolor=three,citecolor=blueone, colorlinks,
   anchorcolor=blue,runcolor=pink,frenchlinks=red
   pdfstartview=FitH,pdftitle=title,%
   pdfauthor=author]{hyperref}

\begin{document}
\title{Description of the newly observed $\Omega^{*}_c$ states as molecular states}

\author{Jingwen Feng}
\affiliation{School of Physics and Electronic Engineering, Sichuan Normal University, Chengdu 610101, China}

\author{Feng Yang}
\affiliation{School of Physical Science and Technology, Southwest Jiaotong University, Chengdu 610031,China}

\author{Cai Cheng}
\email{ccheng@sicnu.edu.cn}
\affiliation{School of Physics and Electronic Engineering, Sichuan Normal University, Chengdu 610101, China}

\author{Yin Huang}
\email{huangy2019@swjtu.edu.cn}
\affiliation{School of Physical Science and Technology, Southwest Jiaotong University, Chengdu 610031,China}

\begin{abstract}
In this work, we study the strong decays of the newly observed $\Omega^{*}_c(3185)$ and $\Omega^{*}_c(3327)$ assuming that $\Omega^{*}_c(3185)$ and $\Omega^{*}_c(3327)$
as $S$-wave $D\Xi$ and $D^{*}\Xi$ molecular state, respectively.  Since the $\Omega_c^{*}$ was observed in the $\Xi_c^{+}K^{-}$ invariant mass distributions,
the partial decay width of $\Omega^{*}_c(3185)$ and $\Omega^{*}_c(3327)$ into $\Xi_c^{+}K^{-}$ through hadronic loops are evaluated with the help of the effective Lagrangians.
Moreover, the decay channel of $\Xi_c^{'}\bar{K}$ is also included.  The decay process is described by the $t$-channel $\Lambda$, $\Sigma$ baryons and $D_s$, $D_s^{*}$
mesons exchanges, respectively.  By comparison with the LHCb observation, the current results support the $\Omega^{*}_c(3327)$ with$J^P=3/2^{-}$ as pure $D^{*}\Xi$ molecule
while  the $\Omega^{*}_c(3327)$ with $J^P=1/2^{-}$ can not be well reproduced in the molecular state picture. In addition, the spin-parity $J^P=1/2^{-}$ $D\Xi$ molecular
assumptions for the $\Omega^{*}_c(3185)$ can't be conclusively determined.  It may be a meson-baryon molecule with a big $D\Xi$ component.  Although the decay width of the  $\Omega_c^{*}\to{}\bar{K}\Xi_c^{'}$ is of the order several MeV, it can be well employed to test the molecule interpretations of $\Omega^{*}_c(3185)$ and $\Omega^{*}_c(3327)$.
\end{abstract}

\date{\today}


\maketitle
\section{Introduction}\label{sec:intro}
After the discovery of the five $\Omega^{*}_c$ states in one observation simultaneously back in 2017~\cite{LHCb:2017uwr}, the latest experimental results
of the LHCb Collaboration reported the existence of an additional two $\Omega_c$ states in the $\Xi_c^{+}K^{-}$ invariant mass spectrum, originating
from $pp$ collisions~\cite{LHCb:2023rtu}.  The measured parameters of these newly states are as follows:
\begin{align}
M_{\Omega_c(3185)}&=3185.1\pm{}1.7^{+7.4}_{-0.9}\pm{}0.2~~~~~{\rm MeV},\nonumber\\
\Gamma_{\Omega_c(3185)}&=50\pm7^{+10}_{-20}~~~~~ {\rm MeV},\\
M_{\Omega_c(3327)}&=3327.1\pm{}1.2^{+0.1}_{-1.3}\pm{}0.2~~~~~{\rm MeV},\nonumber\\
\Gamma_{\Omega_c(3327)}&=20\pm5^{+13}_{-1}~~~~~ {\rm MeV}.\nonumber
\end{align}
Compared to the earlier discovery of two ground states $\Omega^{*}_c(2695)$ and $\Omega^{*}_c(2770)$~\cite{Workman:2022ynf}, the findings presented in not only expand our understanding of the charmed baryon with quantum numbers $C=1$ and $S=2$, which is composed of one charm quark and two strange quarks, but also help us to strange quarks, but also help us to understand the formation mechanism of exotic hadron states. This is mainly because the newly observed $\Omega^{*}_c$ states may display a more complex internal structure than the ground states $\Omega^{*}_c(2695)$ and $\Omega^{*}_c(2770)$, which are made up of three quarks.

Indeed, the $\Omega^{*}_c(3050)$ and $\Omega_c^{*}(3119)$ were suggested to be the exotic pentaquarks in the chiral quark-soliton model~\cite{Kim:2017jpx,Kim:2017khv}.
Similar qualitative results can also be found in Refs.~\cite{Yang:2017rpg,An:2017lwg}.  In Refs.~\cite{Debastiani:2017ewu,Montana:2017kjw},
based on the analysis of the mass spectrum,  the $\Omega^{*}_c(3050)$ was identified as a meson-baryon molecule with $J^P=1/2^{-}$.  This is the same with the results
in Ref.~\cite{Huang:2018wgr} that the total decay widths of the $\Omega^{*}_c(3050)$ can be well reproduced with the assumption that $\Omega^{*}_c(3050)$ as $S$-wave $\Xi{}D$
bound state with $J^P=1/2^{-}$.  By solving a coupled-channel Bethe Salpeter equation using a $SU(6)_{lsf}\times$ HQSS extended WT interaction as a kernel, three
meson-baryon molecular states were obtained~\cite{Nieves:2017jjx} that can be associated to the experimental states $\Omega^{*}_c(3000)$, $\Omega^{*}_c(3050)$, and
$\Omega^{*}_c(31119)$ or $\Omega^{*}_c(3090)$.

There also exist some clues support the newly observed $\Omega^{*}_c(3185)$ and $\Omega^{*}_c(3227)$ as molecular states.
It was suggested in Ref.~\cite{Yu:2023bxn} that $\Omega^{*}_c(3185)$ and $\Omega^{*}_c(3327)$ were proposed to be the $2S(3/2^{+})$ and $1D(3/2^{+})$ states,
respectively.   However, a completely different conclusion was drawn from Ref.~\cite{luo:2023bxn} that the $\Omega^{*}_c(3327)$ is a good candidate of $\Omega^{*}_c(1D)$
state with $J^P=5/2^{+}$.  There exists spin-parity puzzle for $\Omega^{*}_c(3327)$ indicate it difficult to put $\Omega^{*}_c(3327)$ into the conventional $css$ states.
Since the mass of $\Omega^{*}_c(3327)$ is very close to the threshold of $D^{*}\Xi$, the hadronic molecular configuration for $\Omega^{*}_c(3327)$ is possible.  Although
the $\Omega^{*}_c(3185)$ can be considered as a conventional charmed baryon~\cite{Yu:2023bxn},  the hadronic molecule interpretations cannot be excluded due to the mass
of $\Omega^{*}_c(3185)$ is about 6.37 MeV below the threshold of $D\Xi$.

However, no work has been conducted to study whether the newly observed $\Omega^{*}_c(3185)$ and $\Omega^{*}_c(3327)$ can be explained as $D\Xi$ and $D^{*}\Xi$ molecular state,
respectively.  In this work, we estimate possible strong decay modes by assuming $\Omega^{*}_c(3185)$ and $\Omega^{*}_c(3327)$ as molecular states.  By comparing the
calculated decay widths with experimental data, we can evaluate the validity of the proposed molecular explanations for the structures of $\Omega^{*}_c(3185)$ and
$\Omega^{*}_c(3327)$.

This paper is organized as
follows. In Sec.~\ref{Sec: formulism}, we will present the theoretical formalism. In Sec.~\ref{Sec: results}, the numerical result will be given, followed by discussions
and conclusions in last section.
\section{FORMALISM AND INGREDIENTS}\label{Sec: formulism}
\begin{figure}[h!]
\begin{center}
\includegraphics[bb=70 360 1050 710, clip, scale=0.50]{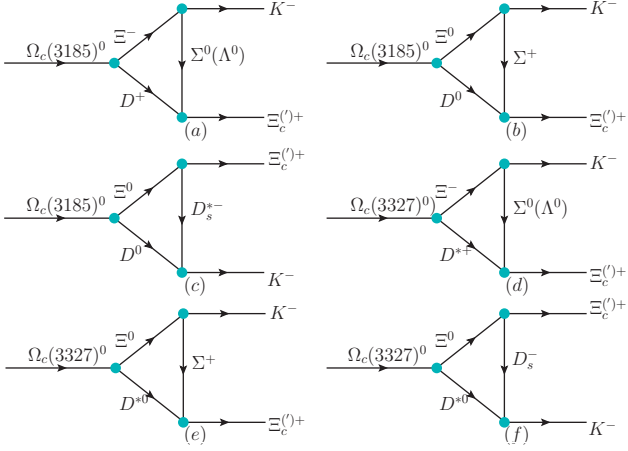}
\caption{Feynman diagrams for the process $\Omega^0_c(3185,3327)\to{}K^{-}\Xi^{+}_c$ and $K^{-}\Xi^{'+}_c$. }\label{cc1}
\end{center}
\end{figure}
With assuming the newly observed $\Omega^{*}_c(3185)$ and $\Omega^{*}_c(3227)$ as $S$-wave molecular states,  the decay of $\Omega^{*}_c(3185)$ and $\Omega^{*}_c(3227)$ into $\bar{K}\Xi_c$
and $\bar{K}\Xi_c^{'}$ are allowed.   The corresponding Feynman diagrams are illustrated in Fig.~\ref{cc1}, which includes the $t$-channel $\Sigma$, $\Lambda$ baryons
exchange and $D_s$, $D_s^{*}$ mesons exchange.  To evaluate these diagrams, the Lagrangian for the coupling between the mesons and the baryons  are obtained using  the $SU(4)$
invariant interaction Lagrangians~\cite{Liu:2001ce}
\begin{align}
\mathcal{L}_{PBB}&=ig_{p}(a\phi^{*\alpha\nu\nu}\gamma^{5}P^{\beta}_{\alpha}\phi_{\beta\mu\nu}+b\phi^{*\alpha\mu\nu}\gamma^{5}P^{\beta}_{\alpha}\phi_{\beta\mu\nu})\label{eq2},\\
\mathcal{L}_{VBB}&=ig_{v}(c\phi^{*\alpha\nu\nu}\gamma P^{\beta}_{\alpha}\phi_{\beta\mu\nu}+d\phi^{*\alpha\nu\nu}\gamma P^{\beta}_{\alpha}\phi_{\beta\mu\nu}),\nonumber
\end{align}
where $V_{\mu}$ and $P$ are the $SU(4)$ vector meson and pseudoscalar meson matrix, respectively.  The meson matrices are
\begin{align}
P=
  \begin{pmatrix}
   \frac{\pi^{0}}{\sqrt{2}}+\frac{\eta}{\sqrt{6}}+\frac{\eta_{c}}{\sqrt{12}} & \pi^{+}&K^{+}&\bar{D}^{0}\\
    \pi^{-} & -\frac{\pi^{0}}{\sqrt{2}}+\frac{\eta}{\sqrt{6}}+\frac{\eta_{c}}{\sqrt{12}}&K^{0}&D^{-}\\
    K^{-}&\bar{K}^{0} & -\frac{2\eta}{\sqrt{6}}+\frac{\eta_{c}}{\sqrt{12}}&D_{s}^{-}\\
    D^{0}&D^{+}&D_{s}^{+}&-\frac{3\eta_{c}}{\sqrt{12}}
\end{pmatrix}
\end{align}
and
\begin{align}
V=
   \begin{pmatrix}
  \frac{\rho^{0}}{\sqrt{2}}+\frac{\omega_{8}}{\sqrt{6}}+\frac{J/\Psi}{\sqrt{12}} & \rho^{+}& K^{*+} & \bar{D}^{*0}\\
    \rho^{-} & -\frac{\rho^{0}}{\sqrt{2}}+\frac{\omega_{8}}{\sqrt{6}}+\frac{J/\Psi}{\sqrt{12}}&K^{*0}&D^{*-}\\
    K^{*-}&\bar{K}^{*0} & -\frac{2\omega_{8}}{\sqrt{6}}+\frac{J/\Psi}{\sqrt{12}}&D_{s}^{*-}\\
    D^{*0}&D^{*+}&D_{s}^{*+}&-\frac{3J/\Psi}{\sqrt{12}}
  \end{pmatrix}.
  \end{align}
with $\omega_8=\omega\cos\theta+\phi\sin\theta$ and $\sin\theta=-0.761$.  The tensors $\phi_{\beta\mu\nu}$ in above equation represents 20-plet of the proton~\cite{Liu:2001ce},
\begin{align}
&p=\phi_{112},n=\phi_{221},\Lambda=\sqrt{\frac{2}{3}}(\phi_{321}-\phi_{312}),\nonumber\\
&\Sigma^{+}=\phi_{113},\Sigma^{0}=\sqrt{2}\phi_{123},\Sigma^{-}=\phi_{233},\nonumber\\
&\Xi^{0}=\phi_{311},\Xi^{-}=\phi_{332},\Sigma^{++}_{c}=\phi_{114},\nonumber\\
&\Sigma^{+}_{c}=\phi_{124},\Sigma^{0}_{c}=\phi_{224},\Xi_{c}^{+}=\phi_{134},\nonumber\\
&\Xi_{c}^{0}=\phi_{234},\Xi_{c}^{+'}=\sqrt{\frac{2}{3}}(\phi_{413}-\phi_{431}),\Xi_{c}^{0'}=\sqrt{\frac{2}{3}}(\phi_{423}-\phi_{432})\nonumber\\
&\Lambda_{c}^{+}=\sqrt{\frac{2}{3}}(\phi_{421}-\phi_{412}),\Omega_{c}^{0}=\phi_{334},\Xi^{++}_{cc}=\phi_{441}\nonumber\\
&\Xi_{cc}^{+}=\phi_{442},\Omega_{cc}^{+}=\phi_{443},
\end{align}
where the indices $\beta$, $\mu$, $\nu$ denote the quark content of the baryon fields with the identification $1\leftrightarrow{}u$, $2\leftrightarrow{}d$,
$3\leftrightarrow{}s$, and $4\leftrightarrow{}c$. Hence, the tensors $\phi_{\beta\mu\nu}$ satisfy following condition
\begin{align}
\phi_{\mu\nu\lambda}+\phi_{\nu\lambda\mu}+\phi_{\lambda\mu\nu}=0,\phi_{\mu\nu\lambda}=\phi_{\nu\mu\lambda}.
\end{align}
By using the exact form of the matrix and the above relationship, the interaction vertices between baryons and pseudoscalar mesons can be estimated by expanding the SU(4) invariant interaction Lagrangians.
The values of the coupling constants adopted in this work can be computed by comparing with the coefficients of the interaction Lagrangians ${\cal{L}}_{\pi{}NN}$ and ${\cal{L}}_{\rho{}NN}$ and determining the constants $g_V$, $g_P$, $a$, $b$, $c$, and $d$ in terms of $g_{\pi{}NN}=13.5$ and $g_{\rho{}NN}=3.25$~\cite{Liu:2001ce}.  Then the values of the coupling constants can be compute and are listed in Tab.~\ref{tab11}.
\begin{table}[h!]
\begin{center}
\caption{Values of the effective couplings constants.}\label{tab11}
\begin{tabular}{cccccccc} 		 	
  \hline\hline
 \toprule
        \textbf{$g_{\Xi_{c}^{+}D^{+}\Sigma^{0}}$} & \textbf{$g_{\Xi^{-}K^{-}\Sigma^{0}}$} & \textbf{$g_{D^{+}\Xi_{c}^{+}\Lambda^{0}}$} & \textbf{$g_{\Xi^{-}K^{-}\Lambda^{0}}$} & \textbf{$g_{D^{0}\Xi_{c}^{+}\Sigma^{+}}$} & \textbf{$g_{\Xi^{0}\Sigma^{+}K^{-}}$} & \textbf{$g_{\Xi^{0}D^{*-}_{s}\Xi_{c}^{+}}$} \\ \midrule
       3.78 & 13.5 & 6.55 & 3.43 & 5.35 & -19.09 & 4.60 \\\midrule
        \textbf{$g_{D^{*+}\Sigma^{0}\Xi_{c}^{+}}$} & \textbf{$g_{D^{*+}\Lambda^{0}\Xi_{c}^{+}}$} & \textbf{$g_{D^{*0}\Sigma^{+}\Xi_{c}^{+}}$} & \textbf{$g_{\Xi^{0}D_{s}^{-}\Xi_{c}^{+}}$} & \textbf{$g_{D^{+}\Sigma^{0}\Xi_{c}^{'+}}$} & \textbf{$g_{D^{+}\Lambda^{0}\Xi_{c}^{'+}}$} & \textbf{$g_{D^{0}\Sigma^{+}\Xi_{c}^{'+}}$} \\ \midrule
       3.25 & -5.63 & -4.60 & 5.35 & 9.48 & 5.47   & 13.41 \\\midrule
        \textbf{$g_{\Xi_{c}^{'+}D^{*-}_{s}\Xi^{0}}$} & \textbf{$g_{D^{*+}\Sigma^{0}\Xi_{c}^{'+}}$} & \textbf{$g_{D^{*+}\Lambda^{0}\Xi_{c}^{'+}}$} & \textbf{$g_{D^{*0}\Sigma^{+}\Xi_{c}^{'+}}$} & \textbf{$g_{\Xi^{0}D_{s}^{-}\Xi_{c}^{'+}}$} & \textbf{$g_{D^{*0}D^{-}_{s}K^{-}}$} & \textbf{$g_{D^{0}D^{*-}_{s}K^{-}}$} \\ \midrule
       -5.63 & 3.98 & -2.30 & 5.63 & -13.40 & 12.00 & 12.00 \\
    \bottomrule
  \hline\hline
  \end{tabular}
 \end{center}
\end{table}

In addition to the Lagrangian shown in Eq.\ref{eq2}, the effective Lagrangians of relevant interaction vertices are also needed~\cite{Yang:2021pio}
\begin{align}
\mathcal{L}_{PPV}&=\frac{iG}{2\sqrt{2}}\langle\partial^{\mu}P(PV_{\mu}-V_{\mu}P)\rangle,\\
\mathcal{L}_{VVP}&=\frac{G'}{\sqrt{2}}\epsilon^{\mu\nu\alpha\beta}\langle\partial_{\mu}V_{\nu}\partial_{\alpha}V_{\beta}P\rangle,\\
\mathcal{L}_{VVV}&=\frac{iG}{2\sqrt{2}}\langle\partial^{\mu}V^{\nu}(V_{\mu}V_{\nu}-V_{\nu}V_{\mu})\rangle,
\end{align}
where the coupling constants $G=12.0$ and $G^{'}=55.51$~\cite{Yang:2021pio}.

Since the $\Omega^{*}_c(3185)$ and $\Omega^{*}_c(3227)$ states are considered as $S$-wave bound states of $D^{(*)}\Xi$, the couplings of $\Omega^{*}_c(3185)$ and $\Omega^{*}_c(3227)$
to their components are written as
\begin{align}
{\cal{L}}^{1/2^{-}}_{\Omega^{*}_c(3185)}&=\sum_{j=\Xi^{-}D^{+},\Xi^{0}D^{0}}C_{j}g^{1/2^{-}}_{\Omega_{c}(3185)\Xi D}\Omega_{c}(3185)(x)\nonumber\\
                                    &\times\int dy\Phi(y^{2})\Xi(x+\omega_{D} y)D(x-\omega_{\Xi} y),\nonumber\\
{\cal{L}}^{1/2^{-}}_{\Omega^{*}_c(3327)}=&\sum_{j=\Xi^{-}D^{*+},\Xi^{0}D^{*0}}C_{j}g^{1/2^{-}}_{\Omega^{*}_c(3327)\Xi D^{*}}\Omega^{*}_c(3327)(x)\gamma^{\mu}\gamma^{5}\nonumber\\
                                     &\times\int dy\Phi(y^{2})\Xi(x+\omega_{D^{*}} y)D^{*}_{\mu}(x-\omega_{\Xi} y),\nonumber\\
{\cal{L}}^{3/2^{-}}_{\Omega^{*}_c(3327)}=&-i\sum_{j=\Xi^{-}D^{*+},\Xi^{0}D^{*0}}C_{j}g^{3/2^{-}}_{\Omega_{c}(3327)\Xi D^{*}}\Omega_{c}(3327)^{\mu}(x)\nonumber\\
                &\int dy\Phi(y^{2})\Xi(x+\omega_{D^{*}} y)D^{*}_{\mu}(x-\omega_{\Xi} y)\label{eq10},
\end{align}
where $\omega_{\Xi}=m_{\Xi}/(m_{\Xi}+m_{D^{(*)}})$ and $\omega_{D^{(*)}}=m_{D^{(*)}}/(m_{\Xi}+m_{D^{(*)}})$ with $m_{\Xi}$ and $m_{D^{(*)}}$ are the masses
of the $\Xi$ and $D^{(*)}$, respectively.  $C_j=1/\sqrt{2}$ is the isospin coefficient, which is calculated from the following isospin assignments for the $\Xi$
and $D^{(*)}$
$$\begin{pmatrix}
  \Xi^{-}\\
\Xi^{0}
\end{pmatrix}\sim
\begin{pmatrix}
-|\frac{1}{2},-\frac{1}{2}\rangle\\
|\frac{1}{2},+\frac{1}{2}\rangle
\end{pmatrix};~~~
\begin{pmatrix}
  D^{(*)+}\\
D^{(*)0}
\end{pmatrix}\sim
\begin{pmatrix}
|\frac{1}{2},+\frac{1}{2}\rangle\\
|\frac{1}{2},-\frac{1}{2}\rangle
\end{pmatrix}.$$
In the above equation, $\Phi(y^{2})$ is the effective correlation function and serves the following two roles:
1) it shows the distribution of the components in the hadronic molecule, 2) it has the same role with the form factor
that can avoid the Feynman diagram's ultraviolet divergence.  Usually, the correlation function can vanish quickly in
the ultraviolet region.  Here we choose the Fourier transformation of the correlation function to have a Gaussian form
\begin{align}
\Phi(-p_E^2)\doteq{}\exp(-p_E^2/\Lambda^2),
\end{align}
where $p_E$ is the Euclidean Jacobi momentum and the parameter $\Lambda$ is taken as a parameter, which will be discussed later.
The coupling constants of $g^{1/2^{-}}_{\Omega^{*}_{c}(3185)\Xi D}$,
$g^{1/2^{-}}_{\Omega^{*}_c(3327)\Xi D^{*}}$, $g^{3/2^{-}}_{\Omega^{*}_c(3327)\Xi{}D^{*}}$ appearing in Eq.~(\ref{eq10}) are given in Eq.~({\ref{eq12}})
\begin{align}
\Sigma^{1/2^{-}}_{\Omega^{*}_c(3185)}(k_{0})=&(g^{1/2^{-}}_{\Omega_{c}\Xi D})^{2}\int_{0}^{\infty} d\alpha \int_{0}^{\infty}d\beta \sum_{j}C_{j}{\cal{Y}}(\omega_{D},m_{D})\nonumber\\
                                         &\times{\cal{Z}}(\omega_{D},m_{D}),\nonumber\\
\Sigma^{1/2^{-}}_{\Omega^{*}_c(3327)}(k_{0})=&(g^{1/2^{-}}_{\Omega_{c}\Xi D^{*}})^{2}\int_{0}^{\infty} d\alpha \int_{0}^{\infty}d\beta \sum_{j}C_{j}{\cal{Y}}(\omega_{D^{*}},m_{D^{*}})\nonumber\\
 &\times{}[2{\cal{Z}}(\omega_{D^{*}},m_{D^{*}})+\frac{k^{3}_{0}(-4\omega_{D^{*}}-2\beta)^{3}}{8m^{2}_{D^{*}}z^{3}}\nonumber\\
                    &-\frac{3k_{0}\Lambda^{2}(-4\omega_{D^{*}}-2\beta)}{2m^{2}_{D^{*}}z^{2}}+\frac{k_{0}^{3}(-4\omega_{D^{*}}-2\beta)}{4m^{2}_{D^{*}}z^{2}}\nonumber\\
                    &-\frac{k^{2}_{0}(-4\omega_{D^{*}}-2\beta)^{2}m_{\Xi}}{4m^{2}_{D^{*}}z^{2}}+\frac{k_{0}\Lambda^{2}}{m^{2}_{D^{*}}z}+\frac{2\Lambda^{2}m_{\Xi}}{m^{2}_{D^{*}}z}],\nonumber
\end{align}
\begin{align}
\Sigma^{T3/2^{-}}_{\Omega^{*}_c(3327)}(k_{0})=&g_{\Omega_{c}\Xi D^{*}}^{2}\int_{0}^{\infty} d\alpha \int_{0}^{\infty}d\beta \sum_{j}C_{j}{\cal{Y}}(\omega_{D^{*}},m_{D^{*}})\nonumber\\
                    &\times [{\cal{Z}}(\omega_{D^{*}},m_{D^{*}})+\frac{k_{0}\Lambda^{2}(-4\omega_{D^{*}}-2\beta)}{4m^{2}_{D^{*}}z^{2}}\nonumber\\
                    &+\frac{k_{0}\Lambda^{2}}{2m^{2}_{D^{*}}z}+\frac{\Lambda^{2}m_{\Xi}}{2m^{2}_{D^{*}}z}]\label{eq12},
\end{align}
with
\begin{align}
{\cal{Y}}(\omega_{D^{(*)}},m_{D^{(*)}})&=\frac{1}{16\pi^2z^2}\exp\{-\frac{1}{\Lambda^{2}}[-2k_{0}^{2}\omega_{D}^{2}+\alpha m_{D}^{2}\nonumber\\
         &+\beta(-k_{0}^{2}+m^{2}_{\Xi})+\frac{(-4\omega_{D}-2\beta)^{2}k_{0}^{2}}{4z}]\}\nonumber\\
{\cal{Z}}(\omega_{D^{(*)}},m_{D^{(*)}})&=k_{0}+m_{\Xi}+\frac{k_{0}(-4\omega_{D}-2\beta)}{2z},
\end{align}
where $Z=2+\alpha+\beta$ and $k_0^2=m^2_{\Omega^{*}_c}$ with $k_0$, $m_{\Omega^{*}_c}$ are the four momenta and the mass of the newly observed $\Omega^{*}_c$, respectively.
If the $\Omega_c^{*}$ is a $D^{(*)}\Xi$ molecular states with $J^P=1/2^{-}$, the $\Sigma^{1/2^{-}}$ is the self-energy operator of the hadronic molecule $\Omega_c^{*}$.
However, $\Sigma^{T3/2^{-}}$ is the transverse part of the self-energy operator $\Sigma^{\mu\nu}_{\Omega_c^{*}}$ when we assuming $\Omega_c^{*}$ as $D^{(*)}\Xi$
molecular states with $J^P=3/2^{-}$.   Once the self-energy operator and its transverse part are obtained, the coupling constants of the hadronic molecule $\Omega_c^*$ to its
constituents $D^{(*)}\Xi$ can be determined using the compositeness condition based on the work in Ref.~\cite{Weinberg:1962hj}.  This condition requires
that the renormalization constant of the hadronic molecular wave function is equal to zero
\begin{align}
&1-\frac{d\Sigma_{\Omega^{*}_c(3185/3327)}}{dk_0}=0,~~~~~J=\frac{1}{2}\nonumber\\
&1-\frac{d\Sigma^{T}_{\Omega^{*}_c(3327)}}{dk_0}=0.~~~~~~~~~~~~~J=\frac{3}{2}\label{eq14}
\end{align}

With the above prepared, we can obtain the general expressions of the amplitudes corresponding to the Feynman diagrams Fig.~\ref{cc1}
\begin{align}
\mathcal{M}_{a}^{1/2^{-}}&=\mu(p_{2})[\frac{1}{\sqrt{2}}g_{\Omega11}g_{\Xi^{(')+}_{c}D^{+}\Sigma^{0}}g_{\Xi^{-}K^{-}\Sigma^{0}}\int \frac{d^{4}k_{1}}{(2\pi)^{4}}\nonumber\\
                         &\times\Phi[(p\omega_{D^{+}}-q\omega_{\Xi^{-}})^{2}]\gamma^{5}\frac{i(k\!\!\!/_{1}+m_{\Sigma^{0}})}{k_{1}^{2}-m_{\Sigma^{0}}^{2}}\gamma^{5}\nonumber\\
                         &\times\frac{i(p\!\!\!/+m_{\Xi^{-}})}{p^{2}-m_{\Xi^{-}}^{2}}\frac{i}{q^{2}-m_{D^{+}}^{2}}+\frac{1}{\sqrt{2}}g_{\Omega11}g_{\Xi^{(')+}_{c}D^{+}\Lambda^{0}}\nonumber\\
                         &\times{}g_{\Xi^{-}K^{-}\Lambda^{0}}\int \frac{d^{4}k_{1}}{(2\pi)^{4}}\Phi[(p\omega_{D^{+}}-q\omega_{\Xi^{-}})^{2}]\gamma^{5}\nonumber\\
                         &\times\frac{i(k\!\!\!/_{1}+m_{\Lambda^{0}})}{k_{1}^{2}-m_{\Lambda^{0}}^{2}}\gamma^{5}\frac{i(p\!\!\!/+m_{\Xi^{-}})}{p^{2}-m_{\Xi^{-}}^{2}}\frac{i}{q^{2}-m_{D^{+}}^{2}}]\mu(k_{0}),\\
\mathcal{M}_{b}^{1/2^{-}}&=\frac{1}{\sqrt{2}}g_{\Omega12}g_{D^{0}\Xi_{c}^{(')+}\Sigma^{+}}g_{\Xi^{0}K^{-}\Sigma^{+}}\int\frac{d^{4}k_{1}}{(2\pi)^{4}}\nonumber\\
                          &\times\phi[(p\omega_{D^{0}}-q\omega_{\Xi^{0}})^{2}]\mu(p_{2})\gamma^{5}\frac{i(k\!\!\!/_{1}+m_{\Sigma^{+}})}{k_{1}^{2}-m_{\Sigma^{+}}^{2}}\gamma^{5}\nonumber\\
                          &\times\frac{i(p\!\!\!/+m_{\Xi^{0}})}{p^{2}-m_{\Xi^{0}}^{2}}\mu(k_{0})\frac{i}{q^{2}-m_{D^{0}}^{2}},
\end{align}
\begin{align}
\mathcal{M}_{c}^{1/2^{-}}&=i\frac{1}{2}g_{\Omega12}g_{D^{0}D^{*-}_{s}K^{-}}g_{\Xi^{0}\Xi_{c}^{(')+}D^{*-}_{s}}\int\frac{d^{4}k_{1}}{(2\pi)^{4}}\nonumber\\
                         &\times\Phi[(p\omega_{D^{0}}-q\omega_{\Xi^{0}})^{2}]\mu({p_{2}})\gamma^{\nu}\frac{i(-g^{\mu\nu}+\frac{k_{1}^{\mu}k_{1}^{\nu}}{m_{D^{*-}_{s}}^{2}})}{k^{2}_{1}-m^{2}_{D^{*-}_{s}}}\nonumber\\
                         &\times(iq^{\mu}-ip_{1}^{\mu})\frac{i}{q^{2}-m_{D^{0}}^{2}}\mu(k_{0})\frac{i(p\!\!\!/+m_{\Xi^{0}})}{p^{2}-m^{2}_{\Xi^{0}}},\\
\mathcal{M}_{d}^{1/2^{-}}&=\mu(p_{2})[\frac{1}{\sqrt{2}}g_{\Omega21}g_{\Xi^{(')+}_{c}D^{*+}\Sigma^{0}}g_{\Xi^{-}K^{-}\Sigma^{0}}\int\frac{d^{4}k_{1}}{(2\pi)^{4}}\nonumber\\
                          &\times\Phi[(p\omega_{D^{*+}}-q\omega_{\Xi^{-}})^{2}]\gamma^{\mu}\frac{i(k\!\!\!/_{1}+m_{\Sigma^{0}})}{k_{1}^{2}-m_{\Sigma^{0}}^{2}}\gamma^{5}\nonumber\\
                          &\times\frac{i(p\!\!\!/+m_{\Xi^{-}})}{p^{2}-m_{\Xi^{-}}^{2}}\gamma^{\nu}\gamma^{5}\frac{i(-g^{\mu\nu}+\frac{q^{\mu}q^{\nu}}{m^{2}_{D^{*+}}})}{q^{2}-m_{D^{*+}}^{2}}\nonumber\\
                          &+\frac{1}{\sqrt{2}}g_{\Omega21}g_{\Xi^{(')+}_{c}D^{*+}\Lambda^{0}}g_{\Xi^{-}K^{-}\Lambda^{0}}\int \frac{d^{4}k_{1}}{(2\pi)^{4}}\nonumber\\
                          &\times\Phi[(p\omega_{D^{*+}}-q\omega_{\Xi^{-}})^{2}]\gamma^{\alpha}\frac{i(k\!\!\!/_{1}+m_{\Lambda^{0}})}{k_{1}^{2}-m_{\Lambda^{0}}^{2}}\gamma^{5}\nonumber\\
                          &\times\frac{i(p\!\!\!/+m_{\Xi^{-}})}{p^{2}-m_{\Xi^{-}}^{2}}\gamma^{\beta}\gamma^{5}\frac{i(-g^{\alpha\beta}+\frac{q^{\alpha}q^{\beta}}{m^{2}_{D^{*+}}})}{q^{2}-m_{D^{*+}}^{2}}]\mu(k_{0}),\\
\mathcal{M}_{d}^{3/2^{-}}&=\mu(p_{2})[-i\frac{1}{\sqrt{2}}g_{\Omega31}g_{\Xi^{(')+}_{c}D^{*+}\Sigma^{0}}g_{\Xi^{-}K^{-}\Sigma^{0}}\int\frac{d^{4}k_{1}}{(2\pi)^{4}}\nonumber\\
                         &\times\Phi[(p\omega_{D^{*+}}-q\omega_{\Xi^{-}})^{2}]\gamma^{\mu}\frac{i({k\!\!\!/_{1}}+m_{\Sigma^{0}})}{k_{1}^{2}-m_{\Sigma^{0}}^{2}}\gamma^{5}\nonumber\\
                         &\times\frac{i({p\!\!\!/}+m_{\Xi^{-}})}{p^{2}-m_{\Xi^{-}}^{2}}g^{\nu\lambda}\frac{i(-g^{\mu\nu}+\frac{q^{\mu}q^{\nu}}{m^{2}_{D^{*+}}})}{q^{2}-m_{D^{*+}}^{2}}\nonumber\\
                         &-i\frac{1}{\sqrt{2}}g_{\Omega31}g_{\Xi^{(')+}_{c}D^{*+}\Lambda^{0}}g_{\Xi^{-}K^{-}\Lambda^{0}}\int \frac{d^{4}k_{1}}{(2\pi)^{4}}\nonumber\\
                         &\times\Phi[(p\omega_{D^{*+}}-q\omega_{\Xi^{-}})^{2}]\gamma^{\alpha}\frac{i({k\!\!\!/_{1}}+m_{\Lambda^{0}})}{k_{1}^{2}-m_{\Lambda^{0}}^{2}}\gamma^{5}\nonumber\\
                         &\times\frac{i({p\!\!\!/}+m_{\Xi^{-}})}{p^{2}-m_{\Xi^{-}}^{2}}g^{\beta\lambda}\frac{i(-g^{\alpha\beta}+\frac{q^{\alpha}q^{\beta}}{m^{2}_{D^{*+}}})}{q^{2}-m_{D^{*+}}^{2}}]\mu_{\lambda}(k_{0}),\\
\mathcal{M}_{e}^{1/2^{-}}&=\frac{1}{\sqrt{2}}g_{\Omega22}g_{\Xi^{(')+}_{c}D^{*0}\Sigma^{0}}g_{\Xi^{0}K^{-}\Sigma^{0}}\int \frac{d^{4}k_{1}}{(2\pi)^{4}}\nonumber\\
                         &\times\Phi[(p\omega_{D^{*0}}-q\omega_{\Xi^{0}})^{2}]\mu(p_{2})\gamma^{\mu}\frac{i({k\!\!\!/_{1}}+m_{\Sigma^{+}})}{k_{1}^{2}-m_{\Sigma^{+}}^{2}}\gamma^{5}\nonumber\\
                         &\times\frac{i({p\!\!\!/}+m_{\Xi^{0}})}{p^{2}-m_{\Xi^{0}}^{2}}\mu(k_{0})\gamma^{\nu}\gamma^{5}\frac{i(-g^{\mu\nu}+\frac{q^{\mu}q^{\nu}}{m^{2}_{D^{*0}}})}{q^{2}-m_{D^{*0}}^{2}},\\
\mathcal{M}_{e}^{3/2^{-}}&=-\frac{i}{\sqrt{2}}g_{\Omega22}g_{\Xi^{(')+}_{c}D^{*0}\Sigma^{0}}g_{\Xi^{0}K^{-}\Sigma^{0}}\int \frac{d^{4}k_{1}}{(2\pi)^{4}}\nonumber\\
                         &\times\Phi[(p\omega_{D^{*0}}-q\omega_{\Xi^{0}})^{2}]\mu(p_{2})\gamma^{\mu}\frac{i({k\!\!\!/_{1}}+m_{\Sigma^{+}})}{k_{1}^{2}-m_{\Sigma^{+}}^{2}}\gamma^{5}\nonumber\\
                         &\times\frac{i({p\!\!\!/}+m_{\Xi^{0}})}{p^{2}-m_{\Xi^{0}}^{2}}\mu^{\nu}(k_{0})\frac{i(-g^{\mu\nu}+\frac{q^{\mu}q^{\nu}}{m^{2}_{D^{*0}}})}{q^{2}-m_{D^{*0}}^{2}},
\end{align}
\begin{align}
\mathcal{M}_{f}^{1/2^{-}}&=\frac{i}{2}g_{\Omega31}g_{D^{*0}D^{-}_{s}K^{-}}g_{\Xi^{0}D_{s}^{-}\Xi_{c}^{(')+}}\int \frac{d^{4}k_{1}}{(2\pi)^{4}}\nonumber\\
                         &\times\Phi[(p\omega_{D^{*0}}-q\omega_{\Xi^{0}})^{2}]\mu(p_{2})\gamma^{5}\frac{1}{k_1^{2}-m_{D_{s}^{-}}^{2}}\nonumber\\
                         &\times(p_{1}^{\mu}-k_{1}^{\mu})\frac{(-g^{\mu\nu}+\frac{q^{\mu}q^{\nu}}{m^{2}_{D^{*0}}})}{q^{2}-m_{D^{*0}}^{2}}\mu(k_{0})\gamma^{\nu}\nonumber\\
                         &\times\gamma^{5}\frac{({p\!\!\!/}+m_{\Xi^{0}})}{p^{2}-m_{\Xi^{0}}^{2}},\\
\mathcal{M}_{f}^{3/2^{-}}&=\frac{1}{2}g_{\Omega32}g_{D^{*0}D^{-}_{s}K^{-}}g_{\Xi^{0}D_{s}^{-}\Xi_{c}^{(')+}}\int \frac{d^{4}k_{1}}{(2\pi)^{4}}\nonumber\\
                          &\Phi[(p\omega_{D^{*0}}-q\omega_{\Xi^{0}})^{2}]\mu(p_{2})\gamma^{5}\frac{1}{k_1^{2}-m_{D_{s}^{-}}^{2}}\nonumber\\
                          &\times(p_{1}^{\mu}-k_{1}^{\mu})\frac{(-g^{\mu\nu}+\frac{q^{\mu}q^{\nu}}{m^{2}_{D^{*0}}})}{q^{2}-m_{D^{*0}}^{2}}\mu^{\nu}(k_{0})\frac{({p\!\!\!/}+m_{\Xi^{0}})}{p^{2}-m_{\Xi^{0}}^{2}},
\end{align}
where the $p_1$, $p_2$, $p$, $q$, and $k_1$ are the four momenta of the $K$, $\Xi_c^{(')}$, $\Xi$, $D^{(*)}$, and $t$-channel exchanged particles, respectively.

Once the amplitudes are determined, the corresponding decay widths can be obtained, which read,
\begin{align}
\Gamma(\Xi_c^{*}\to)=\frac{1}{2J+1}\frac{1}{8\pi}\frac{|\mathbf{p}_1|}{m_{\Omega_c^{*2}}}\overline{{\cal{M}}^2},
\end{align}
where $J$ is the total angular momentum of the initial  state $\Omega_c^{*}$, the overline indicates the sum over the polarization
vectors of final hadrons.  Here $\mathbf{p}_1$ is the 3-momenta of the decay products in the center of mass frame.

\section{RESULTS AND DISCUSSIONS}\label{Sec: results}
In this work, we study the strong decay pattern of $S$-wave $D^{(*)}\Xi$ molecular states within the effective Lagrangians approach,
and find the relation between the $D^{(*)}\Xi$ molecular state and the newly observed $\Omega^{*}_c(3185)$ and $\Omega^{*}_c(3327)$ states.
If the estimated strong decay width matches well with the LHCb observation, we can judge the molecule explanations for the structure
of $\Omega^{*}_c(3185)$ and $\Omega^{*}_c(3327)$.   To make a reliable prediction for the strong decay width, the parameter $\Lambda$ that affects
our accurate calculation must be clarified.   Since the $\Lambda$ could not be well determined from first principles, it has to be determined
by fitting to the experimental data.  We especially hope the value of the $\Lambda$ can be determined within same theoretical framework adopted
in this work.

Fortunately, within the same theoretical framework adopted in current work in Ref~\cite{Dong:2017rmg}, the parameter $\Lambda$
was constrained as $\Lambda=0.91-1.00$ GeV by comparing the sum of the partial decay modes of the $\eta(2225)$ and $\phi(2170)$ with the total width.
We also note that many exotic states can be well considered as molecules with $\Lambda=0.90-1.10$ GeV, and we refer the reader to review the
Refs.~\cite{Yang:2021pio, Dong:2008gb, Huang:2020taj} and their reference.  Therefore, we take $\Lambda$ in the range of $\Lambda=0.90-1.10$ GeV
to study whether the $\Omega^{*}_c(3185)$ and $\Omega^{*}_c(3327)$ can be interpreted as molecule composed of $D^{(*)}\Xi$ molecular components.

\begin{figure}[h!]
\begin{center}
\includegraphics[bb=-25 192 850 350, clip, scale=0.95]{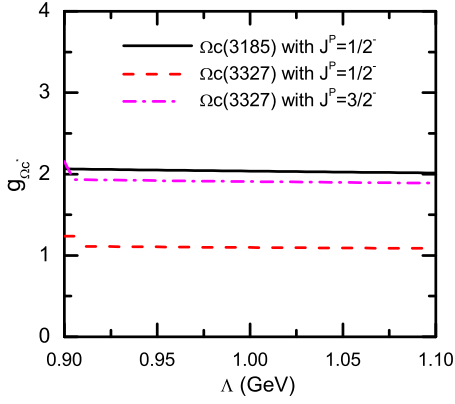}
\caption{Coupling constants for the newly observed $\Omega_c^{*}$ with different spin-parity as a function of the parameter $\Lambda$.}\label{cc2}
\end{center}
\end{figure}
Considering the $\Lambda$ values adopted in this work,  we will first discuss the results of the calculative coupling constants.  Substituting
Eq.~(\ref{eq12}) into Eq.~(\ref{eq14}), the dependence of the coupling constants on the parameter $\Lambda$ is solved.  By performing the
integration of parameters $\alpha$ and $\beta$ from $0$ to infinite, the numerical results are presented in Fig.~\ref{cc2}, which shows the variation of
the coupling constants with respect to the $\Lambda$ in the range of 0.90 to 1.10 GeV.
From the results, we can find that the value of the coupling constants decreases with the increase of $\Lambda$, and they are not very sensitive to the $\Lambda$.
It is important to note that the coupling constant $g_{\Omega^{*}_c(3327)\Xi{}D^{*}}$ decreases sharply at a small $\Lambda$ value.  Due to the presence of ultraviolet
divergences in the calculation, the average value of the molecular mass was used.  Here, the divergences originate from the threshold of $D^{0}\Xi^0$ and $D^{*0}\Xi^0$
channels being slightly below the masses of the $\Omega^{*}_c(3185)$ and $\Omega^{*}_c(3327)$, respectively.

\begin{figure}[h!]
\begin{center}
\includegraphics[bb=8 240 850 420, clip, scale=0.95]{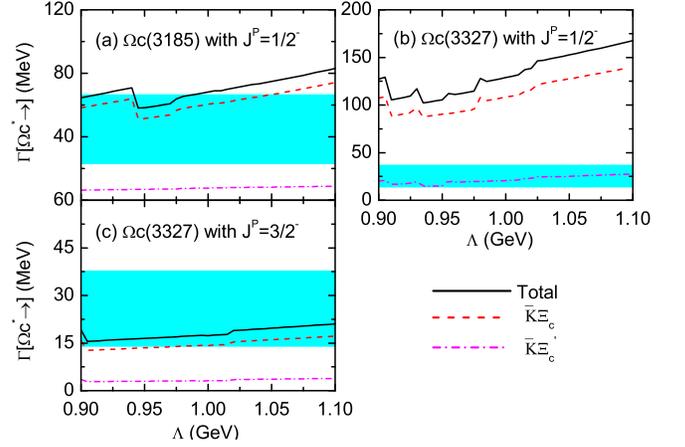}
\caption{Partial decay widths of the $\Omega_c^{*}\to{}\bar{K}\Xi_c$ (red dash line), $\Omega_c^{*}\to{}\bar{K}\Xi_c^{'}$ (magenta dash dot line), and the total decay width with different $\Omega_c^{*}$ states depending on the parameter $\Lambda$.   The oycn error bands correspond to the LHCb observed~\cite{LHCb:2023rtu}.}\label{cc3}
\end{center}
\end{figure}

After obtaining the coupling constants, we compute the partial decay widths for the transitions $\Omega^{*}_c(3185)\to{}\bar{K}\Xi^{(')}_c$ and $\Omega^{*}_c(3327)\to\bar{K}\Xi_c^{(')}$,
and plot the results in Fig.~\ref{cc3}, which vary with the parameter $\Lambda=0.9-1.1$ GeV.  Note that the partial decay widths in the channels $\Omega^{*}_c(3185)\to{}K^{-}\Xi^{(')+}_c$ $\Omega^{*}_c(3185)\to{}K^{-}\Xi^{(')+}_c$ and $\Omega^{*}_c(3327)\to{}K^{-}\Xi_c^{(')+}$ are only estimated here, and the other channels  $\Omega^{*}_c(3185)\to{}\bar{K}^0\Xi^{(')0}_c$ and $\Omega^{*}_c(3327)\to{}\bar{K}^{0}\Xi_c^{(')0}$ can be obtained by isospin symmetry.  The sum of these partial decay widths gives the total decay width of the $\Omega^{*}_c(3185)$ or total decay width of the $\Omega^{*}_c(3185)$ or $\Omega^{*}_c(3327)$, which are also shown in Fig.~\ref{cc3}.

From the results of Fig.~\ref{cc3}(a), we can find that the total decay width for the transitions $\Omega^{*}_c(3185)\to{}\bar{K}\Xi^{(')}_c$
increases with the increase of $\Lambda$, while it decreases when $\Lambda$ varies from 0.940 to 0.945 GeV.  Comparing with the total decay
width for the $\Omega^{*}_c(3185)\to{K}\Xi_c^{(')}$ reactions, we found the line shape for the total decay width of the $\Omega^{*}_c(3327)\to{}\bar{K}\Xi_c^{'}$
reactions are very different.  To see how different, we take $\Omega^{*}_c(3327)$ at $J^P=1/2^{-}$ case as an example.  The obtained total decay
width for the $\Omega^{*}_c(3327)\to{}\bar{K}\Xi_c^{'}$ reaction decreases, then it begin increases at $\Lambda=0.935$ GeV.

Now, let me give a clear discussion about whether the $\Omega^{*}_c(3185)$ and $\Omega^{*}_c(3327)$ can be interpreted as molecules composed of $D^{(*)}\Xi$
molecular components.  From Fig.~\ref{cc3}, we observe that the total decay width for the $\Omega^{J^P=1/2^{-}}_c(3327)\to \bar{K}\Xi_c^{(')}$
and $\Omega^{J^P=3/2^{-}}_c(3327)\to \bar{K}\Xi_c^{(')}$ is estimated to be about 104.39-167.54 MeV and 15.55-21.04 MeV, respectively, where the
theoretical value 15.55-21.04 MeV is close to the experimental data $20\pm5^{+13}_{-1}$.  This suggests if the spin-parity of the $\Omega^{*}_c(3327)$
is $J^P=3/2^{-}$, the assignment as an $S$-wave pure $D^{*}\Xi$ molecular state for the $\Omega^{*}_c(3327)$ is supported.  However, if the $\Omega^{*}_c(3327)$
has a spin-parity of $1/2^{-}$, the predicted total decay width is much bigger than the experimental total width, which disfavors such a spin-parity
assignment for the $\Omega^{*}_c(3327)$ in the$D^{*}\Xi$ molecular picture.  Moreover, the spin-parity $J^P=1/2^{-}$ $D\Xi$ molecular assumptions for the
$\Omega^{*}_c(3185)$ cannot be conclusively determined.  This is because the obtained total decay width for the $\Omega^{*}_c(3185)$ with $J^P=1/2^{-}$ $D\Xi$
assignment is comparable with that of the experimental total width in the range of $\Lambda=0.900-0.915$ and $\Lambda=0.945-0.990$ GeV.  It may be a
meson-baryon molecule contain a big $D\Xi$ component.

Indeed, there exist several coupled states composed of $D\Xi$ and $D^{*}\Xi$ components~\cite{Zhu:2022fyb}.  However, the authors in Ref.~\cite{Zhu:2022fyb}
claim that the interaction between $D^{(*)}$ meson and $\Xi$ baryon is not strong enough to form a pure bound state with quantum numbers considered in the
current work.  A possible reason for this is that the potential kernels they obtained only consider light meson exchanges and do not include the contact term
that describes the short distance interaction between the $\Xi$ and charmed meson.

Fig.~\ref{cc3} also tells us that the decay width of $\Omega^{*}_c(3185)$ into $\bar{K}\Xi_c$ is about 64.60-82.98 MeV, which almost fully accounts for the total width
of $\Omega_c^{*}(3185)$.  In other words, the transition from $\Omega^{*}_c(3185)$ to $\bar{K}\Xi_c$, which is the experimental observation channel,  provides a
dominant contribution to the total decay width.  However, the decay width of $\Omega_c^{*}(3185)\to \bar{K}\Xi_c$ is up to several MeV, which accounts for
9.86\%-10.70\% of its total width.  That means the transition $\Omega_c^{*}(3185)\to\bar{K}\Xi^{'}_c$ gives a minor contribution.  A possible explanation for this
may be that the phase space for  the transition $\Omega_c^{*}(3185)\to\bar{K}\Xi^{'}_c$ is smaller than that of the $\Omega_c^{*}(3185)\to\bar{K}\Xi_c$ reaction.
The same conclusion can also be drawn for the $\Omega_c^{}(3327)\to\bar{K}\Xi_c^{(')}$ reactions.

It should be noted that the authors in Ref.~\cite{Yu:2023bxn} only compute the decay widths of $\Omega_c^{*}(3185)$ and $\Omega_c^{*}(33327)$ into $\bar{K}\Xi_c$ by
assigning $\Omega_c^{*}(3185)$ and $\Omega_c^{*}(3327)$ as state $2S(3/2^{+})$ and $1D(3/2^{+})$, respectively.  The predicted decay widths can reach 78.16 MeV and
25.60 MeV, respectively, which are both larger than the experimental cental value.   This suggest that if the newly observed $\Omega_c^{*}$ is conventional three quark
state,  the transition $\Omega_c^{*}\to\bar{K}\Xi^{'}_c$ is very small and can be ignored.   Thus, we propose the experimental search for $\Omega_c^{*}$ in the $\Omega_c^{*}\to\bar{K}\Xi^{'}_c$ reaction that offers a nice channel to test the molecular nature of the $\Omega_c^{*}(3185)$ and $\Omega_c^{*}(33327)$.

\section{Summary}\label{sec:summary}
Inspired by the newly observed baryons $\Omega_c^{*}(3185)$ and $\Omega_c^{*}(33327)$,  the strong decay widths of the $\Omega_c^{*}\to{}\bar{K}\Xi_c$ and $\Omega_c^{*}\to{}\bar{K}\Xi_c^{'}$ was studied in an effective Lagrangian approach.  Our theoretical approach is based on the assuming that the newly observed
$\Omega^{*}_c(3185)$ and $\Omega^{*}_c(3327)$ can be explained as $S$-wave $D\Xi$ and $D^{*}\Xi$ molecular state, respectively.  With the $D^{(*)}\Xi$ assignment,
the partial decay widths of the $\Omega_c^{*}(3185)$ and $\Omega_c^{*}(33327)$ into the $\bar{K}\Xi_c^{(')}$ final states through hadronic loop are calculated.
The decay process is described by the $t$-channel $\Lambda$, $\Sigma$ baryons and $D_s$, $D_s^{*}$ mesons exchanges, respectively.

The current results support the $\Omega^{*}_c(3327)$ with$J^P=3/2^{-}$ as pure $D^{*}\Xi$ molecule
while  the $\Omega^{*}_c(3327)$ with $J^P=1/2^{-}$ can not be well reproduced in the molecular state picture.  In addition, the spin-parity $J^P=1/2^{-}$ $D\Xi$ molecular
assumptions for the $\Omega^{*}_c(3185)$ can't be conclusively determined.  It may be a meson-baryon molecule with a big $D\Xi$ component.  We also find that the transition $\Omega^{*}_c(3185)\to\bar{K}\Xi_c$, which is the experimental observation channel,  provides a dominant contribution to the total decay width while the transition $\Omega_c^{*}(3185)\to\bar{K}\Xi^{'}_c$ gives a minor contribution.  However, the decay channel $\Omega_c^{*}\to{}\bar{K}\Xi_c^{'}$ can be well employed to test the molecule interpretations of $\Omega^{*}_c(3185)$ and $\Omega^{*}_c(3327)$ by
comparing our results with this in Ref.~\cite{Yu:2023bxn}.

\section*{Acknowledgments}
Yin Huang acknowledges the YST Program of the APCTP and the support from the National Natural Science Foundation
of China under Grant No.12005177.


\end{document}